\documentclass[12pt]{iopart}
\usepackage{iopams,graphicx}
\begin{document}

\title{Constraints on non-thermal Dark Matter from {\sc Planck} lensing extraction}
 
\author{L.A. Popa  and A. Vasile
\footnote[3]{To whom correspondence should be addressed: lpopa@venus.nipne.ro} }

\address{ISS Institute for Space Sciences  
Bucharest-Magurele, Ro-76900 Romania}

\eads{\mailto{lpopa@venus.nipne.ro} and 
\mailto{avasile@venus.nipne.ro}}

\date{\today} 

\begin{abstract} 
Distortions of CMB temperature and polarization anisotropy maps caused by gravitational lensing,
observable with high angular resolution and sensitivity, can be used to constrain the
sterile neutrino mass,  offering
several advantages against the analysis based on the combination of CMB, LSS and Ly$\alpha$ forest power spectra. As the gravitational lensing effect depends on the matter distribution, no 
assumption on light-to-mass bias is required. In addition, unlike the galaxy clustering 
and Ly$\alpha$ forest power spectra, the projected gravitational potential power spectrum probes a larger range of angular scales, the non-linear corrections being required only at very small scales. \\
Taking into account the changes in the time-temperature relation of the primordial plasma and
the modification of the neutrino thermal potential, we  compute the projected gravitational 
potential power spectrum and its correlation with the temperature in 
the presence of DM sterile neutrino. \\
We show that the cosmological parameters are generally not biased when DM sterile neutrino is included. 
From this analysis we found a lower limit on DM sterile neutrino mass $m_{\nu_s}>2.08$ keV at 95\% CL, 
consistent with the lower mass limit obtained from the combined analysis of CMB, SDSS 3D power spectrum and SDSS Ly$\alpha$ forest power spectrum ($m_{\nu_s}>1.7$ keV).\\
We conclude that although the information that can be obtained
from  lensing extraction is rather limited due to the high level of the lensing noise of  {\sc Planck} experiment, weak lensing of CMB offers a valuable alternative 
to constrain the dark matter sterile neutrino mass. 
\end{abstract}

\pacs{CMBR theory, dark matter, cosmological neutrinos}

\maketitle

\section*{Introduction}

The latest results of the cosmic microwave background (CMB) temperature and polarization 
anisotropies from WMAP 3-year observations \cite{Nolta,Page} combined with other tracers 
such as large-scale structure (LSS) galaxy surveys, supernovae luminosity distance 
and Ly$\alpha$ forest, have lead in specifying the $\Lambda$CDM model as the cosmological concordance model \cite{Spergel06,Seljak07}. 

The direct confirmation of the $\Lambda$CDM theory is the detection of the acoustic Doppler peaks
structure of the CMB angular power spectrum. Further successes are related to the correct
prediction of the hierarchical structure formation via gravitational instability,
the abundance of clusters at small redshifts, the spatial distribution and the number
density of galaxies, the LSS matter power spectrum, the Ly$\alpha$  forest
amplitude and spectrum \cite{Turner00}. 
Despite its successes on large scales, the $\Lambda$CDM model produces too  much
power on small scales. In general, the observed structures have softer cores, lower concentrations and are less clumped than those predicted by the
$\Lambda$CDM model (see Ref. \cite{Prim01} and references therein). 

A possibility to alleviate the accumulating contradiction between the $\Lambda$CDM model
predictions on small scales and the observations is to add properties to the dark matter (DM) sector,
relaxing the hypothesis on dark matter as being cold. 
Free streaming due to the thermal motion  of the DM particles is the simplest known 
mechanism for smearing out small scale structure. 
Between the large free streaming distance of Hot Dark Matter (HDM) particles 
and the small free streaming distance of Cold Dark Matter (CDM) particles lies the intermediate scale 
of Warm Dark Matter (WDM) particles \cite{Colombi,Bode}. 
The velocity dispersion of WDM particles is sufficient to alleviate some contradictions between the $\Lambda$CDM model predictions and observations
such as: the predicted number of halos compared with the observed number of satellite galaxies in the Local Group \cite{Bode,Padm95}, the prediction of  $1/r^{\alpha}$ ($1 < \alpha < 1.2$) behavior for galaxy rotation curves
compared with the observed linearly rising behavior \cite{Dal,Bosch,Zen}, the prediction of
too much baryonic material with low angular momentum to form the observed rotationally disk galaxies \cite{Dolgov01}. 

On the other hand, in addition to the evidence for mixing of active neutrinos from solar and atmospheric oscillation experiments, there are indications for another oscillation with  larger mass-squared 
difference coming from short base-line neutrino oscillation experiments  
\cite{Atha,Miniboone} that can be explained 
by adding one or two sterile neutrinos with eV-scale mass to the standard scheme with three 
active neutrinos (see Ref.\cite{Maltoni} for a recent analysis).\\
Such results have impact on cosmology because sterile neutrinos can contribute 
to the number of relativistic degrees of freedom at the Big Bang nucleosynthesis \cite{Cirelli}. 
These models are subject to strong bounds on the sum of active neutrino masses 
from the combination of various cosmological data sets \cite{Rafelt,Find} that  
rule out a thermalized sterile neutrino component with eV mass-scale \cite{Seljak07,Mel}. 

However, there is the possibility to accommodate the cosmological observations 
with data from short base-line neutrino oscillation experiments by postulating 
that  sterile neutrino is not thermalized and has a phase-space distribution 
significantly suppressed relative to the thermal distribution.     
It is already known that sterile neutrino with  mass of few keV 
- the Tremaine-Gunn \cite{Tremaine} bound - provides a 
valuable DM candidate \cite{DodWidr,Dol,Aba01,Sha06}. 
In the standard non-resonant production mechanism 
(small lepton number  of the order of baryon number, $L \sim 10^{-10}$) sterile neutrinos are produced 
via small mixing angle oscillation conversion of thermal active neutrinos \cite{Aba02}.  

Sterile neutrino with the mass in keV range  has a radiative decay channel 
emitting a photon with an energy that is half of its mass eigenstate. 
The width of the decay line  increases as the fifth power of its mass eigenstate and 
as square of its mixing angle,  being potentially detectable in various X-ray spectra 
of the astrophysical objects \cite{Pal,Tuck}.
The strategy to search for dark matter particles possesing a radiative decay channel and to derive 
the constraints on their parameters from  X-ray observations applied to sterile neutrino   
is discussed in Ref.\cite{Boy06,Riemer}.\\
The analysis of the observed X-ray background from HEAO-1 and XMM-Newton \cite{Boy1} 
leads to an upper mass limit for sterile neutrino of $m_{\nu_s} < 8.9$ keV 
\footnote{Throughout the paper the sterile neutrino mass is quoted at 95\% CL.} improved 
to $m_{\nu_s} < 6.3$ keV
from the combined analysis of XMM-Newton observations of the Virgo and Coma clusters \cite{Boy2}.      
Also, the constraints on the rate  
of sterile neutrino radiative decay obtained from 
the analysis of the diffuse X-ray spectrum of 
Andromeda galaxy leads to $m_{\nu_s} < 3.5$ keV,  
which is a significant improvement over previous upper limits \cite{Watson}.\\
Although prone to uncertainties due to the estimate of DM distribution in 
dwarf galaxy \cite{AbaKou}, interesting constraints on the parameters of radiative DM decay  
are obtained from XMM-Newton observations of the
 DM halo of the Milky Way and Ursa Minor \cite{Boy3}.
The recent spectral analysis of the unresolved component of the cosmic X-Ray background in 
the {\sc Chandra} North and South Deep Fields  provides limits on the sterile neutrino mass \cite{Aba07}.
The highest Milky Way halo mass estimate provides a 
limit on sterile neutrino mass of $m_{\nu_s} < 2.9$ keV 
in Dodelson-Widrow production model \cite{DodWidr}, while the lowest halo mass estimates provides a more conservative limit of $m_{\nu_s} < 5.7$ keV.

The radiative decay of the sterile neutrino can also augment the 
ionization fraction of the primordial gas at high redshifts 
leading to the increase of the temperature of the primordial gas, the enhancement  molecular hydrogen formation and  of the star formation rate \cite{Kus06,Mapelli,Pierpa}.

The combined analysis of the Ly$\alpha$ forest   
power spectrum measured by SDSS, the CMB anisotropy and the galaxy 
clustering power spectra yielded 
to the lower limits for sterile neutrino mass, $m_{\nu_s} >14$ keV \cite{Sel06}  
and $m_{\nu_s} > 10$ keV \cite{Viel06} in Dodelson-Widrow production model \cite{DodWidr},  
excluding sterile neutrino as DM candidate. 
However, the method based on the combined analysis of angular power spectra 
gives direct limits for the free streaming lengths of DM particles, 
the limits on their masses 
depending on their momentum distribution functions and therefore 
on their production mechanisms.

Taking into account the deviations from the thermal spectrum of DM particles produced due to 
the variation of the number of degrees of freedom leading to  changes in the time-temperature relation
of the primordial plasma and the modification of the neutrino thermal potential 
\cite{Aba01, Aba02, AbaPrd}, the combination of CMB, LSS and Ly$\alpha$ forest
angular power spectra leads to  $m_{\nu_s} > 1.7$ keV with a further improvement to $m_{\nu_s} > 3$ keV  when
high-resolution Ly$\alpha$ forest power spectra  are considered \cite{AbaPrd}.

In this paper we explore an alternative method to constrain the sterile neutrino   
mass based on the CMB weak lensing extraction. 
Distortions of CMB temperature and polarization maps caused by the gravitational 
lensing potential, observable with high angular resolution and sensitivity, 
have impact on cosmological parameter degeneracies when non-minimal cosmological scenarios 
are considered \cite{AbaDod,Song,Julien,SelZal}. \\
The CMB weak lensing offers several advantages against 
the method based on the combination of angular power spectra. 
As the gravitational lensing effect depends on the dark matter distribution in the Universe, 
no assumption on light-to-mass bias is required.
The projected gravitational potential
is sensitive to the matter distribution out to high redshifts, 
preventing from non-linear corrections required only at very small scales.   
In addition, unlike the galaxy clustering and the Ly$\alpha$ forest,
the projected gravitational potential probes a larger range of angular scales, 
most of the signature comming from large scales.

The paper is organized as follows. In Section~1 we compute the energy density evolution 
of the dark matter sterile neutrino in the expanding Universe, 
incorporating the time-temperature relation 
and the modification of the neutrino thermal potential.
In Section~2 we compute the deflection angle power spectrum and its cross-corelation 
with the temperature when sterile neutrino dark matter energy density is considered.
In Section~3 we derive limits on the sterile neutrino mass from {\sc Planck} weak lensing 
extraction. We draw our main conclusions in Section~4.    

Throughout we assume a background cosmology consistent
with the most recent cosmological measurements \cite{Spergel06} with
energy density of $\Omega_m=0.3$ in matter, $\Omega_b=0.05$ in baryons,
$\Omega_{\nu}=0.01$ in three active neutrino flavors,
$\Omega_{\Lambda}=0.7$ in cosmological constant, a Hubble constant of
$H_0$=72 km s$^{-1}$Mpc$^{-1}$, a reionization redshift $z_{re}$=13 (assuming 
a sharp reionization history) and  adiabatic initial conditions with
a power-law scalar spectral index $n_s=1$.

\section{Sterile neutrino dark matter production and density evolution}

\subsection{Non-resonant production}

Sterile neutrinos dark matter candidates are produced in the early Universe through 
 oscillation with active neutrinos, at temperatures close to QCD phase transition 
\cite{Aba01,Aba02}. 
The oscillation process takes place due to the fact that 
the neutrino mass eigenstate components propagate differently  
as they have different energies, momenta and masses. 
Eigenstates of neutrino interaction include the active neutrinos, $\nu_a$ 
(a=$e$, $\mu$, $\tau$), which are created and destroyed in
the standard model by weak interactions as well as the  sterile neutrinos, $\nu_s$, 
which do not participate in weak interactions.  
For oscillations occurring in vacuum between two neutrino flavors $\nu_{a}$ and
$\nu_s$, the mixing can be written as:
\begin{eqnarray}
\vert   \nu_{a}>=\cos\theta_0 \vert \nu_1 > + \sin \theta_0 \vert \nu_2> \nonumber \\
\vert   \nu_s>=-\sin \theta_0 \vert \nu_1 > + \cos \theta_0 \vert \nu_2>\, ,
\end{eqnarray}
where $\nu_1$ and $\nu_2$ are the neutrino mass eigenstate components and $\theta_0$
is the vacuum mixing angle. The mixing of antineutrinos can be obtained from the above equation by performing the transformation $\nu_a \rightarrow {\bar \nu}_a$
and $\nu_s \rightarrow {\bar \nu}_s$. 
It is usual to consider that each neutrino/antineutrino of a definite flavor is dominantly one mass eigenstate. In this circumstance we refer to the dominant mass 
eigenstate component of $\nu_a/{\bar \nu}_a$ as $\nu_1$  to that of $\nu_s/{\bar \nu}_s$ as 
$\nu_2$ and to their difference of squared masses as 
$\delta m^2= m^2_{\nu_s}-m^2_{\nu_a}$.

The mixing of the mass eigenstate components are modified in the presence of the 
finite temperature background either in the case of small active neutrino lepton number, $L_{\nu_a}\sim10^{-10}$ (of the order of baryon number), driving non-rezonant sterile neutrino production \cite {DodWidr}, 
or in the case of several orders of magnitude larger lepton number \cite {Aba01,shi99}.
The lepton number is defined as the difference between the active
neutrino and antineutrino number densities normalized by the photon number density: 
$L_{\nu_a}=(n_{\nu_a}-n_{{\bar \nu}_a})/n_{\gamma}$.   

The matter mixing angle, $\theta_M$, is related to the vacuum mixing angle, $\theta_0$, through (see e.g.   
Ref. \cite{Aba01} and references therein):
\begin{equation}
\sin^2 2\theta_M
=\frac{\Delta^2(p) \sin^2 2\theta_0}
{\Delta^2(p) \sin^2 2\theta+ D^2(p)+
[ \Delta(p)\cos 2\theta+V^L-V^T(p) ] ^2}. 
\end{equation}
In the above equation $\Delta(p)=\delta m^2/2p$ is the vacuum oscillation factor, $V^L$ is the asymmetric lepton potential, $V^T(p)$ is the thermal potential, $D(p)=\Gamma_{\nu_a}(p)/2$ is the quantum damping rate and $\Gamma_{\nu_a}$ is the collision rate defined as:
\begin{displaymath}
\Gamma_{\nu_a}(p,T)\approx 
\left\{
\begin{array}{ll}
1.27 G^2_F p T^4, & a=e\\
0.92 G^2_F p T^4, & a=\mu,\tau 
\end{array} \right. 
\end{displaymath}
The asymmetric lepton potential, $V^L$, and the thermal potential 
$V^T(p)$ read as \cite{Aba01,Aba02}:
\begin{eqnarray}
V^L=\sqrt{2}G_F \left[ 2 ( n_{\nu_a}-n_{{\bar \nu}_a}) 
+\sum_{ a \neq a^{'}  } 
( n_{\nu_{a^{'}}}-n_{{\bar \nu}_{a^{'}}})-\frac{n_n}{2}\right]\,,& \nonumber \\
V^T=-\frac{8\sqrt{2}G_Fp}{3m^2_Z}( <E_{\nu_a}>n_{\nu_a}+
<E_{\bar{\nu}_a}>n_{{\bar \nu}_a}) \nonumber \\
-\frac{8\sqrt{2}G_Fp}{3m^2_W}( <E_{\nu_a}>n_{\nu_a}+
<E_{\bar{\nu}_a}>n_{{\bar \nu}_a}) \,,   
\end{eqnarray} 
where: $E_{\nu_a}$ is the active neutrino total energy,  $n_{\nu_a}/n_{\bar{\nu}_a}$ are neutrino/antineutrino number densities,cosmological 
$n_n$ is the baryon number density, 
$m_Z$ is $Z^0$ boson mass and  $m_W$ is $W^{\pm}$ boson mass. 
The number density of the thermally distributed active neutrinos, $n_{\nu_a}$ is given by:
\begin{equation}
n_{\nu_a}=\frac{3}{8}n_{\gamma} \left(   \frac{T_{\nu_a}}{T_{\gamma}}\right)^3\,, 
\end{equation}
where $n_{\gamma}=2 \zeta(3)T_{\gamma}^3/\pi^2$ is the photon number density and $\zeta(3)$ is the Riemann zeta function of 3.\\
Throughout we will consider a lepton symmetric Universe 
($n_{\nu_a}\simeq n_{{\bar \nu}_a}$) and neglect the contribution to the asymmetric potential due to 
the baryon number which is very small at the temperatures of interest \cite{Aba02}. 
In this case the contribution of asymmetric lepton potential $V^L$ in eq. (2) is negligible. 

\subsection{Energy density evolution}

\begin{figure}
\begin{center}
\includegraphics[height=10cm,width=14cm]{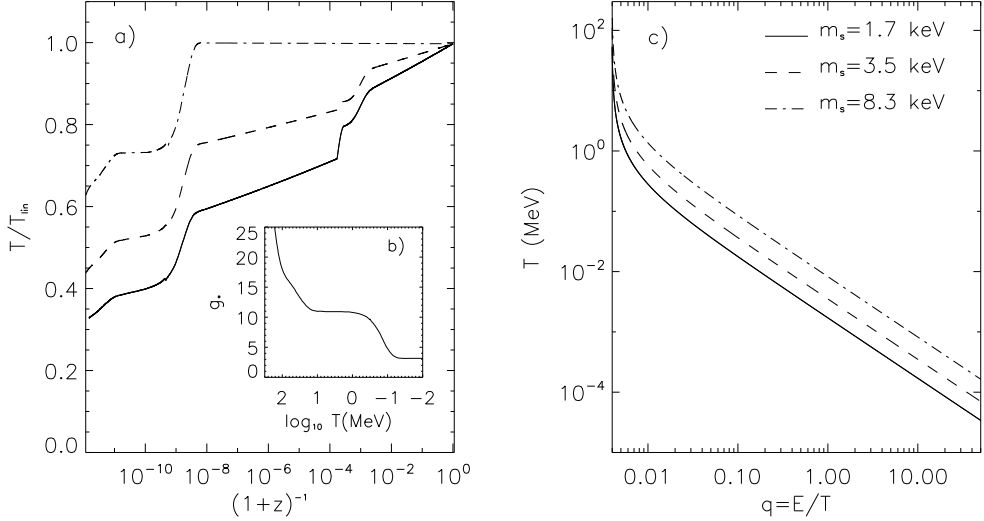}
\caption{Panel a): The evolution with the scale factor of the temperature 
obtained by the integration of the time-temperature relation 
considering the variation of the statistical weight in relativistic particles $g^*$, 
compared with the linear evolution of the temperature, $T_{lin}=T_{CMB}(1+z)$.
Panel b): The evolution with the temperature of the statistical weight in relativistic 
particles $g^*$ used in the computation of the time-temperature relation.
Panel~c): The temperature dependence on  sterile neutrino comoving  momentum obtained for the same sterile neutrino mass values as in Panel~a).}
\label{fig1}
\end{center}
\end{figure}
Since sterile neutrinos/antineutrinos are produced non-thermally, 
their mean energy density and pressure must be computed by 
the direct integration of their  phase-space distributions:
\begin{eqnarray}
\rho_{\nu_s}+\rho_{\bar{\nu}_s}=\frac{T^4}{2 \pi^2}\int_0^{\infty} dq \, q^2E_{\nu_s}
[f_{\nu_s}(q)+f_{{\bar\nu}_s}(q)] \,, \nonumber \\
p_{\nu_s}+p_{{\bar \nu_{s} }}=\frac{T^4}{2 \pi^2}\int_0^{\infty} dq \frac{q^2} {E_{\nu_s}}
[f_{\nu_s}(q)+f_{{\bar\nu}_s}(q)]  \,,
\end{eqnarray}   
where: $f_{\nu_s}$ and $f_{{\bar\nu}_s}$
are the neutrino/antineutrino phase-space distribution functions and  $E_{\nu_s}=\sqrt{q^2+a^2m^2_{\nu_s}}$ 
is the  total energy of sterile neutrino having a mass $m_{\nu_s}$ and a comoving momentum 
$q=E/T$; $a=(1+z)^{-1}$ is the scale factor ($a_0$=1 today). 
The temporal evolution of sterile neutrino energy density affects the 
Hubble expansion rate that reads as:  
\begin{equation}
H(a)=\sqrt{\frac{8 \pi G}{3}}[\Omega_m/a^3+\Omega_r/a^4+\Omega_{\Lambda}]^{1/2}\,. 
\end{equation}
In the above equation $G$ is the gravitational constant, $\Omega_m=\Omega_b+\Omega_{cdm}+\Omega_{\nu_a}+\Omega_{\nu_s}$
is the matter energy density parameter where
$\Omega_b$, $\Omega_{cdm}$, $\Omega_{\nu_a}$ and $\Omega_{\nu_s}$ are the energy 
density parameters for baryons, cold dark matter, active  and sterile
neutrinos respectively, $\Omega_r$ is the radiation energy density parameter and 
$\Omega_{\Lambda}$ is the vacuum energy density parameter. 

\begin{figure}
\begin{center}
\includegraphics[height=10cm,width=14cm]{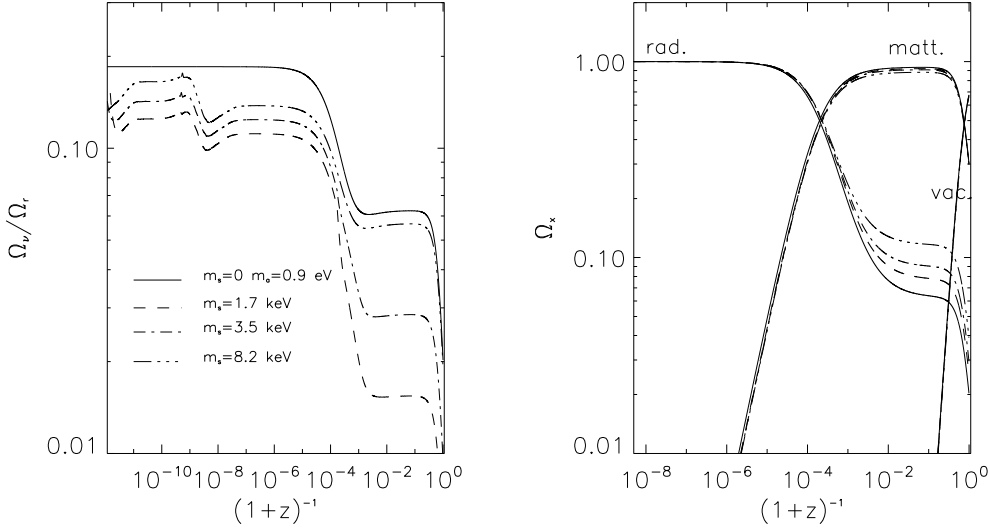}
\caption{ Left: The dependence on the scale factor of the energy density parameters 
of sterile neutrinos expressed in units of the mean energy density of one mass eigenstate of massless neutrinos  for different sterile neutrino mass values. 
Right: The evolution with the scale factor of the energy density parameters for matter, 
radiation and vacuum obtained for fiducial model when 
the corresponding sterile neutrino energy density parameters are included.}
\label{fig1}
\end{center}
\end{figure}
Depending on the mass, the maximum rate production of sterile neutrino occurs  at \cite{DodWidr,Barbi}:
\begin{eqnarray}
T_{max}\approx 133 \, {\rm MeV} \left( \frac{m_{\nu_s}}{1 {\rm keV}}\right)^{1/3}\,.\nonumber 
\end{eqnarray}  
For sterile neutrino with mass in the range of interest, 
the time evolution of the temperature-dependent thermal potential 
and of the collision rate require the 
knowledge of the time-temperature relation in the expanding 
Universe:
\begin{equation}
\frac{da}{dT}=\frac{d\rho_{tot}}{dT}(\rho_{tot}+p_{tot})^{-1}\,,
\end{equation}
where $\rho_{tot}$ and $p_{tot}$ are the total density and pressure, including 
the contribution of sterile neutrino/antineutrino as given in equation (5). 
The time-temperature relation takes  into account 
the variation  of statistical weight in relativistic particles 
that is changed by nearly an order of magnitude 
since $T_{max}$  \cite{Aba01, AbaPrd}.
A general  treatment of the time-temperature relation, 
whose approach we have incorporated, is given in the Appendix of Ref. \cite{Aba01}. \\
Left panel of Figure~1 presents the evolution with the scale factor of the temperature 
obtained by the intergration of the time-temperature relation taking into account
the  variation of the statistical weight in relativistic particles $g^*$,
compared with the linear evolution of the temperature $T_{lin}=T_{CMB}(1+z)$. 
We also show, in the right panel of the same figure, the dependence of the temperature 
on sterile neutrino comoving momentum. 

The time evolution of the sterile neutrino phase-space distribution, $f_{\nu_s}$, can be described by the semiclasical Boltzmann equation \cite{DodWidr,Aba01}: 
\begin{eqnarray}
\frac{\partial }{\partial t}f_{\nu_s}(q)-H(t)\,q\,\frac{\partial}{\partial q}f_{\nu_s}(q) 
\approx \Gamma(\nu_a \rightleftharpoons \nu_s)[f_{\nu_a}(q)-f_{\nu_s}(q)]\,, 
\end{eqnarray}   
where: $dt=da/aH(a)$, $f_{\nu_a}$ is the active neutrino phase-space distribution and $\Gamma(\nu_a \rightleftharpoons \nu_s)$ is the effective production/annihilation rate of sterile neutrinos:
\begin{eqnarray}
 \Gamma(\nu_a \rightleftharpoons \nu_s)\approx 
0.25 \Gamma_{\nu_a}(p,T) \sin^2 2 \theta_M . 
\end{eqnarray}
We use the set of equations (5)-(9) to compute the temporal evolution of the energy density
of sterile neutrinos in the expanding Universe.
For sterile neutrino with the mass in the range of interest, 
the flavor dependence is almost negligible \cite{Aba02}, implying that similar results 
can be obtained for all  $\nu_a \rightleftharpoons \nu_s$ mixings. 
Throughout we consider $\nu_{\tau} \rightleftharpoons \nu_s$ mixing.  \\
Left panel of Figure~2 presents the dependence on the scale factor of the energy density parameter 
of sterile neutrino expressed in units of the mean energy density of one mass eigenstate of massless neutrinos for three different mass values. We also show the same dependence for the case $m_{\nu_s}=0$. 
The right panel of the same figure shows the evolution with the scale factor of the energy density parameters for matter, radiation and vacuum obtained for our flat 
$\Lambda$CDM background cosmology when the corresponding sterile neutrino masses are included.
The production of DM sterile neutrino affects the radiation and matter evolution through 
the Hubble expansion rate, altering the time-temperature relation and redshifting the relativistic species. 

\section{CMB lensing extraction with {\sc Planck} experiment}

\subsection{Deflection angle power spectrum}

The CMB photons from the last scattering surface are subject to the cumulative effects 
of the large scale structure gravitational potential \cite{Hu00}.
The net result is that the CMB temperature anisotropy and polarization patterns measured in the direction ${\bf n}$ in the sky provides information on the photons emerging from the last scattering surface from the direction ${\bf{\tilde n}}={\bf n}+\delta {\bf n}$. The intensity and the linear polarization of the lensed CMB are completely specified by the Stokes parameters $I$, $Q$ and $U$ which are related with the unlensed Stokes parameters through: $X({\bf n})= {{\tilde X}( {\bf{\tilde n}}})$, where $X({\bf n})$ stands for the lensed $I$, $Q$ and $U$. The deflection field is defined as the gradient of the lensing potential 
$ \delta{\bf n}={\bf \nabla} \Psi({\bf n})$, where ${\bf \nabla}$ is the covariant derivative on the sphere.\\
The deflection map and the lesnsing potential map can be expanded in spherical harmonics \cite{Hirata,Cha}, so that the relation between 
the deflection angle power spectrum $C^{dd}_l$ and the lensing 
potential power spectrum $C^{\Psi \Psi}_l$ is given by \cite{Hu02}:
\begin{equation}
C^{dd}_l=l(l+1)C^{\Psi \Psi}_l\,.
\end{equation}

The power spectra of the lensing potential, $C^{\Psi \Psi}_l$ and the correlation with the temperature anisotropy,  $C^{\Psi T}_l$, can be numeriacally computed by using the Code for Anisotropies in the Microwave Background (CAMB) \cite{Hu00,Cha,Lewis} in the linear theory as well
as by including the non-linear corrections from HALOFIT \cite{Smith}. 
We recall that $C^{\Psi T}_l$  does not vanish because the temperature map includes information on 
the time variation of the gravitational potential though the integrated Sachs-Wolfe effect. \\
We modified the CMB anisotropy code CAMB to compute the  CMB angular power spectra in 
the presence of a DM sterile neutrino component. 
We include in the computation the momentum-dependent sterile neutrino phase-space distribution function, as described in the previous section, its unperturbed and perturbed energy density and pressure, energy flux and shear stress.

\begin{figure}
\begin{center}
\includegraphics[height=9cm,width=12cm]{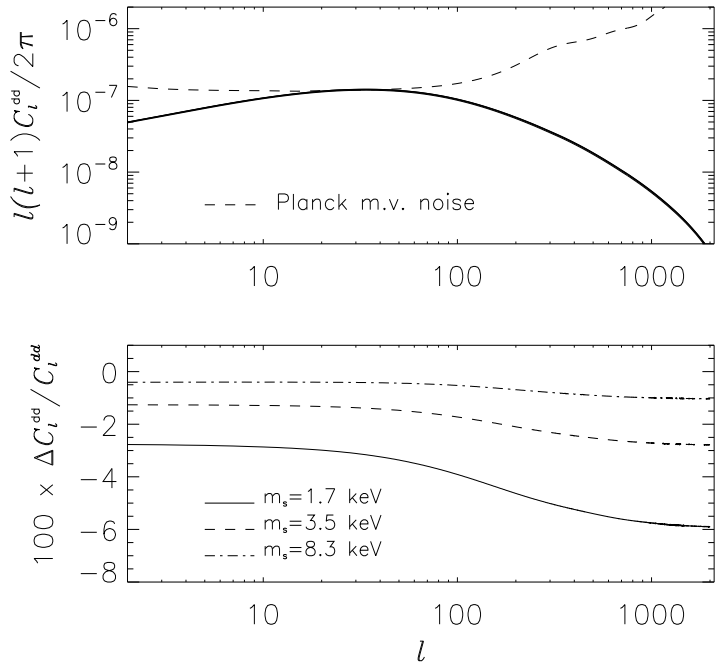}
\caption{Top panel: The deflection angle power spectrum, $C^{dd}_l$, obtained 
in the our fiducial model ($m_{\nu_s}=0$) and the minimum variance noise power spectrum (dashed line) 
for {\sc Planck} experimental characteristics given in Table~1. 
Bottom panel: The relative percentual differences between the fiducial model ($m_{\nu_s}$=0) and exactly the same model with $m_{\nu_s}$=~1.7, 3.5 and 8.3 keV. }
\end{center}
\end{figure}
Figure~3 presents in the top panel the deflection angle power spectrum, $C^{dd}_l$, obtained 
in our fiducial model ($m{\nu_s}=0$). The bottom panel of Figure~3 shows the relative percentual differences between the fiducial model and exact the same model but with $m_{\nu_s}$= 1.7, 3.5 and 8.3 keV.\\
The signature of sterile neutrino mass on the deflection angle power spectrum is the suppression
of its amplitude relative to the fiducial model. The net suppression 
is scale dependent and the relevant length scale is
the free-streaming scale \cite{Bond}, which increases with the mean 
dark matter sterile neutrino velocity and decreases with its mass. 

\section{Constraints on sterile neutrino mass from {\sc Planck} 
lensing extraction}

\subsection{Gaussian Likelihood function}
\begin{figure}
\begin{center}
\includegraphics[height=8cm,width=12cm]{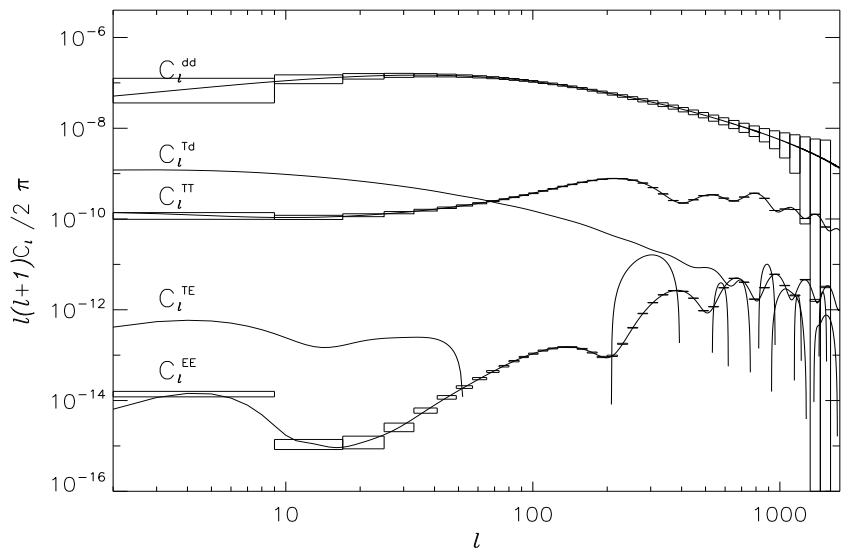}
\caption{ The CMB angular power spectra for the fiducial model.
The error bars include the Planck experiment instrumental noise and the cosmic variance.  
For graphical representation convenience, the error bars assume a multipole binning 
of $\Delta l=7$ until $l\sim 70$ and then $\Delta l\sim l/10$ \cite{Blue,Perotto1}. }
\end{center}
\end{figure}

In the standard inflationary cosmology, the spherical harmonic coefficients of the unlensed 
temperature and polarization field in the sky obey the the Gaussian statistics and can be used 
to define the estimators of the covariance:
\begin{eqnarray} 
 {\hat C^{ab}_l=\frac{1}{2l+1} \sum_m}W^*_{lm}X_{lm} \hspace{0.3cm}  a,b \in \{T, E, B \} \,,
\end{eqnarray}
where $W_{lm}$ and $X_{lm}$ are the harmonic coefficients of the CMB temperature and polarization anisotropy maps. 
The corresponding  full-sky likelihood function is given by:
\begin{equation} 
\chi^2 =-2 \ln {\cal L} ( {\bf \hat C}| {\bf C} ) =
(2l+1)\{ {\rm Tr} [{\bf \hat C} {\bf C^{-1}}] +\ln|{\bf C}| \} \,,
\end{equation}
where {\bf C}  is the covariance matrix of the observations and ${\bf \hat C}$ is the corresponding covariance matrix of estimators.
The covariance matrix of the observations reads as:
\begin{eqnarray}
{\bf C}=\left(\begin{array}{ccc}
{\tilde C}^{TT}_l+N^{TT}_l  &     {\tilde C}^{TE}_l       &     0 \\ 
       {\tilde C}^{TE}_l    & {\tilde C}^{EE}_l+N^{EE}_l  &      0    \\
       0    &        0           & {\tilde C}^{BB}_l+N^{BB}_l
\end{array}\right)  \,. 
\end{eqnarray} 
In the above equation  ${\tilde C}^{ab}_l$, $a,b=\{ T,E, B\}$ are the unlensed power spectra 
of primary anisotropies, $N^{ab}_l$ are the corresponding instrumental noise variances and 
the parity invariance ($C^{BT}_l=C^{BE}_l=0$) it is assumed. 

Weak lensing correlates the lensed multipoles and  the lensed sky is not any more Gaussian \cite{Seljak96}. 
However, as shown in Ref.\cite{Lewis05}, at least up to  {\sc Planck} resolutions and  sensitivities, equation (11) remains approximately correct  if the B-mode, that is noise dominated, is not included in the covariance matrix.  
The non-Gaussian corrections can be however important for the future high sensitivity CMB polarization experiments having a signal dominated B-mode power spectrum \cite{Kapli}. \\
In this paper we consider a fiducial model with no significant amplitude of the primordial
gravitational waves, omitting the B-mode from the parameter estimation analysis. 
In this case, the data covariance matrix read as: 
 \begin{eqnarray}
{\bf C}=\left(\begin{array}{ccc}
C^{TT}_l+N^{TT}_l  &     C^{TE}_l       &     C^{Td}_l \\
       C^{TE}_l    & C^{EE}_l+N^{EE}_l  &      0    \\
       C^{Td}_l    &        0           & C^{dd}_l+N^{dd}_l
\end{array}\right)  \,, 
\end{eqnarray}  
where  $C^{ab}_l$, $a,b=\{ T,E \}$, are the lensed CMB power spectra,  $N^{ab}_l$ are the corresponding detector noise power spectra, $C^{dd}_l$ is deflection angle power spectrum, $N^{dd}_l$ the noise power spectrum associated to the lensing extraction and $C^{Td}_l$ is the power spectrum of the cross-correlation between the temperature and deflection angle.  \\
The weak lensing does not change the total variance in the CMB temperature anisotropy and polarization maps and hence  lensed observations have the same variance as
if there were no lensing \cite{Lewis05,Kapli}.

\begin{table}
\begin{center}
\caption{The expected  experimental characteristics for the {\sc Planck} frequency channels considered in this work \cite{Blue}: $\nu$ is the frequency of the channel, $\theta_b$ is the FWHM,
$\Delta_T$ and $\Delta_P$ are the sensitivities per pixel for temperature and polarization
maps.}
\end{center}
\vspace{0.3cm}
\begin{center}
\begin{tabular}{cccc}
\hline \hline
 $\nu $ & FWHM  & $\Delta_T  $ & $\Delta_P $ \\
 (GHz)&(arc-minutes)&    ($\mu$ K)&       ($\mu$ K)                           \\
\hline \hline
               100 & 9.5 &6.8 & 10.9 \\
               143 & 7.1 &6.0 & 11.4 \\
               217 & 5.0& 13.1 & 26.7 \\
\hline \hline
\end{tabular}
\end{center}
\end{table}
For each frequency channel we consider an homogeneous detector noise with the power spectrum, $N^{ab}_{l,\nu}$, given by \cite{Knox}:
\begin{eqnarray}
N^{ab}_{l,\nu}=(\theta_b \Delta_c)^2 \exp^{ l(l+1) \theta^2_b / 8 \ln 2}\,,
\end{eqnarray}
where $\nu$ is the frequency of the channel, $\theta_b$ is the FWHM of the  beam and
$\Delta_c$, where $c \in(T,P)$, are the sensitivities per pixel for the temperature and polarization maps. The global noise of the experiment, $N^{XX}_l$, can be written as:
\begin{equation}
N^{ab}_l=\left[ \sum_{\nu} (N^{ab}_{l,\nu})^{-1} \right]^{-1} \, .
\end{equation}
The expected experimental performances of the {\sc Planck} frequency channels considered in this paper \cite{Blue} are presented in Table~1. 

The lensing noise power spectrum, $N^{dd}_l$, was numerically computed by using the minimum variance quadratic estimator of Okamoto and Hu \cite{Oka03}. 
By definition, the quadratic estimator is build from pairs 
$(a,b)$ of observed temperature  or polarization modes and its multipoles are given in the quadratic form:
\begin{equation} 
d(a,b)^M_L=N^{ab}_L \sum_{ll{'}mm^{'}}
{{\cal G}(a,b)^{mm^{'}}_{l \,\,\, l^{'}}}^M_L  
a^m_l b^{m^{'}}_{l^{'}} \,,
\end{equation}
where $a^m_l$ and $b^m_l$ correspond to the lensed CMB fields. The minimum variance estimator is obtained by finding the weights ${\cal G}^{a,b}_{l_{1}l_{2}}(L)$ that minimize the Gaussian variance:
\begin{equation}
<{d(a,b)}^{M^{*}}_{L}{d(a^{'},b^{'}})^{M}_{L}>\equiv \delta_{L,L^{'}}\delta_{M,M^{'}}[C^{dd}_L +{\cal N}^{aba^{'}b^{'}}_L]\,.
\end{equation}
For the minimum variance estimator the lensing noise reads as:
\begin{eqnarray}
N^{dd}_{l}=\left[ \sum_{aba^{'}b^{'}} ({\cal N}^{aba^{'}b^{'}})^{-1} \right]^{-1} \, .
\end{eqnarray}
In the top panel of Figure~3 we show the lensing noise power spectrum, $N^{dd}_l$, 
for {\sc Planck} experiment, obtained by using the minimum variance estimator method.\\
Figure~4 presents the CMB angular power spectra  for the fiducial model. 
The error bars, $\Delta C^{\alpha}_l$, include the instrumental noise and the cosmic variance:
\begin{eqnarray}
\hspace{-1cm}\Delta C^{\alpha}_l=\sqrt{  \frac{2}{ (2l+1) \Delta l f_{sky}} }  
( C^{\alpha}_l+ N^{\alpha}_l)\, 
\hspace{0.3cm}\alpha=[TT,EE,TE,dd,Td] \,.&
\end{eqnarray} 
In the above equation $f_{sky}$ represents the fraction of the sky covered by the observations.  For the purpose of this work we take $f_{sky}$=0.8 and assume a perfect cleaning of all the astrophysical foregrounds.  For graphical representation convenience, the error bars presented in Figure~4 assume a multipole binning of $\Delta l=7$ until $l\sim 70$ and then $\Delta l\sim l/10$ \cite{Blue,Perotto1}:
 
\subsection{Parameters estimation}

\begin{figure}
\begin{tabular}{cc}
\includegraphics[height=7cm,width=7cm]{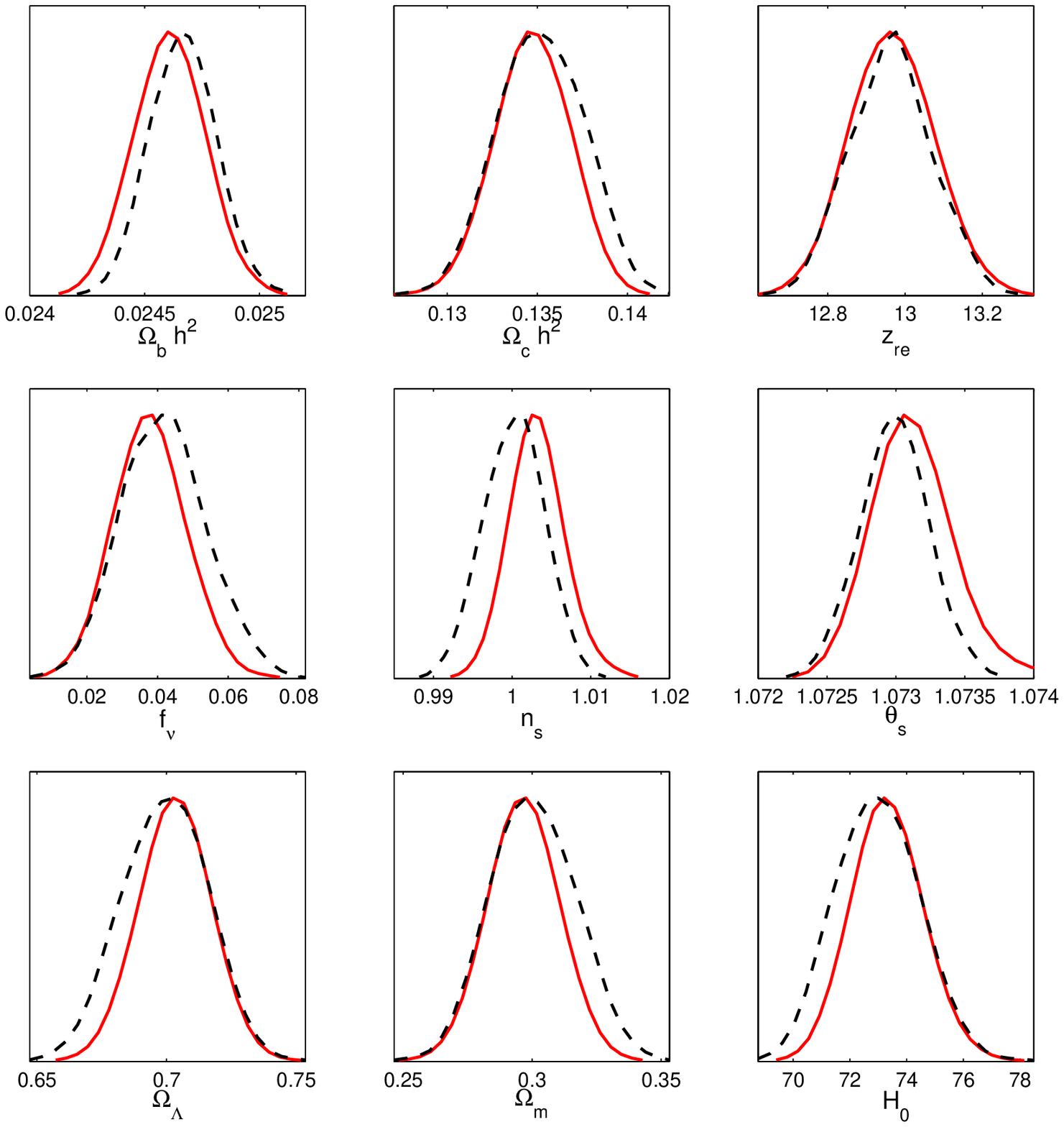}
\includegraphics[height=7cm,width=7cm]{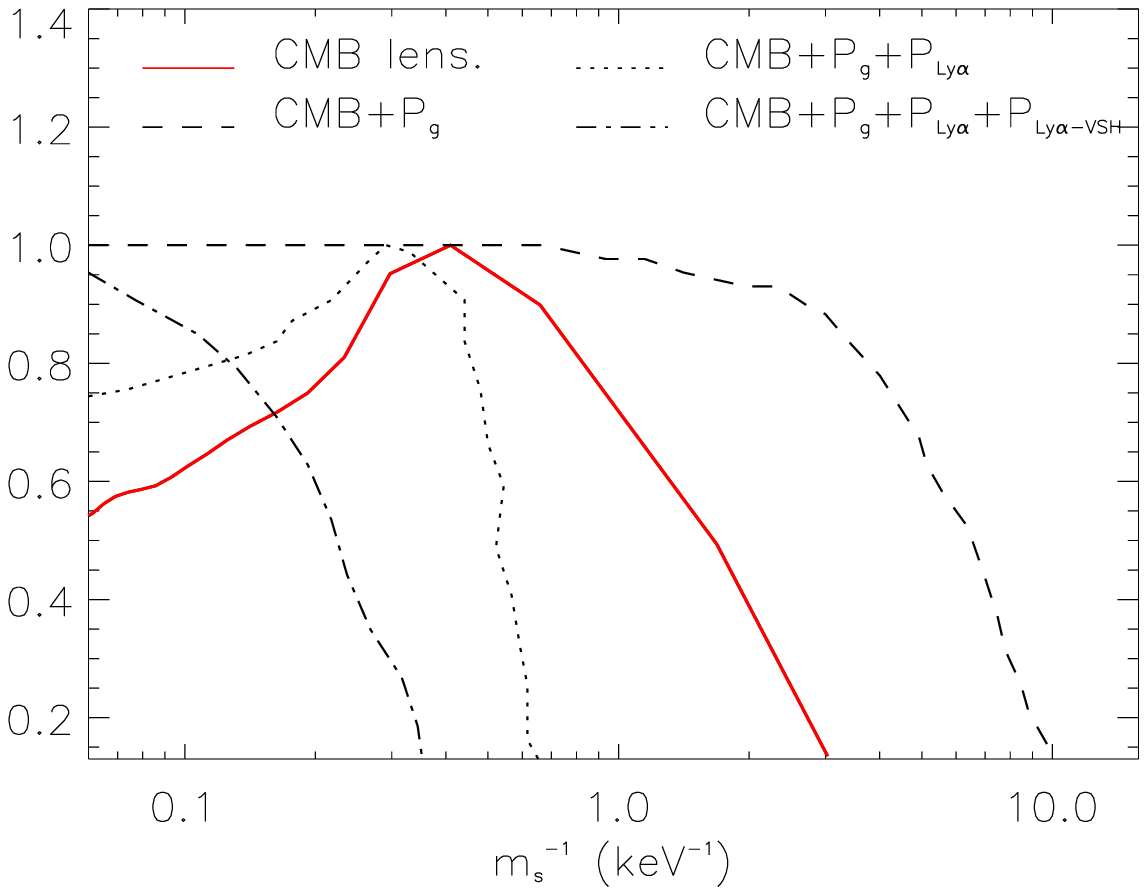} 
\end{tabular}
\caption{Left: The marginalized likelihood probabilities for {\sc Planck}-like 
observations obtained by using the lensing extraction 
with (solid red line) and without (dashed black line) DM sterile neutrino component. 
Right: The marginalized likelihood probabilities for the (inverse) mass of DM 
sterile neutrino obtained for the {\sc Planck}-like 
observations with lensing extraction (solid red line) compared with the  marginalized likelihoods obtained from \cite{AbaPrd}: CMB plus SDSS 3D power spectrum (dashed line), plus SDSS Ly$\alpha$ forest power spectrum (dotted line), plus  VHS high-resolution $Ly{\alpha}$ forest power spectrum (dash-dotted line).}
\end{figure}
We assume a flat $\Lambda$CDM cosmological model with the following set of parameters to 
be determined from the {\sc Planck}-like observations:
\begin{eqnarray}
{\bf \Theta}=(\Omega_bh^2,\Omega_ch^2, \theta_s, z_{re}, f_{\nu},m_{\nu_s}, n_s) \, \nonumber
\end{eqnarray} 
where $\Omega_b$ and $\Omega_c$ are the fractions of critical density in baryons and cold dark matter, $\theta_s$ is the angular acoustic peak scale of the CMB, $z_{re}$ is the  reionization redshift 
(we assume a sharp reionization history), $f_{\nu}$ is the fraction of massive neutrinos, $m_{\nu_s}$ is the
sterile neutrino mass and $n_s$ is the spectral index of primordial adiabatic perturbations. 

We use the {\sc CosmoMC} \cite{Bridle}  Monte Carlo Markov Chain (MCMC) public package
 with the modification of the function {\sc ChiSqExact} 
\cite{Perotto1} required to compute the likelihood function when the deflection angle 
power spectrum is included, 
to sample for the posterior distribution of parameters space ${\bf \Theta}$, 
giving the {\sc Planck}-like data.
For graphical representation convenience, the error bars in Figure~4 assume a multipole binning of $\Delta l=7$ until $l\sim 70$ and then $\Delta l\sim l/10$ \cite{Blue,Perotto1}:
 
Left panel from Figure~5 presents  the marginalized likelihood probabilities 
of the cosmological parameters for  {\sc Planck}-like observations obtained from lensing extraction 
with and without the DM sterile neutrino component, showing that the cosmological parameters 
are generally not biased when including DM sterile neutrino. \\
In the right panel of Figure~5 we present the marginalized likelihood probability for 
the inverse mass, $m^{-1}_s$, of sterile neutrino obtained for  {\sc Planck}-like 
observations by using the lensing extraction. 
We find a lower limit on the sterile neutrino dark matter $m_{\nu_s}>2.08$ keV at 95\% CL, in agreement 
with our previous result \cite{popa} obtained from {\sc Planck} lensing extraction by using the Fisher information matrix ($m_{\nu_s}>1.75$ keV). \\
For comparison we also show in the same figure the marginalized likelihoods 
for $m^{-1}_{\nu_s}$ obtained in Ref. \cite{AbaPrd} from: CMB plus SDSS 3D power spectrum, plus SDSS Ly$\alpha$ forest power spectrum, plus  VHS high-resolution Ly${\alpha}$ forest power spectrum.
Our result is in agreement with the result  from the combined 
analysis of CMB, SDSS 3D power spectrum and SDSS Ly$\alpha$ forest power spectrum ($m_{\nu_s} > 1.7$ keV) 
but is less constrained than the limit obtained when VHS high-resolution Ly$\alpha$ forest data are included ($m_{\nu_s} > 3$ keV).

\section{Conclusions}

From the future high precision  observations of the {\sc Planck} satellite 
we would like to 
extract information about both unlensed  and lensed CMB anisotropies. 
The weak lensing effects of the CMB induced by the neighbouring galaxy clustering 
will have a significant effect on the statistics of the observed CMB 
and must be take into account in order to obtain reliable parameter estimates, offering
several advantages against the cosmological parameter extraction based on the combination of CMB,
 LSS and Ly$\alpha$ power spectra. 
As the gravitational lensing effect depends on the matter distribution, no 
assumption on light-to-mass bias is required. In addition, 
unlike the galaxy clustering and Ly$\alpha$ forest power spectra, 
the projected gravitational potential power spectrum probes 
a larger range of angular scales, the non-linear corrections being 
required only on very small scales.

However, the amount of information that can be obtained from the CMB weak lensing extraction 
depends on the lensing noise level. \\
In this paper we address the possibility to constrain 
the DM sterile neutrino mass from {\sc Planck}-like observations by using the CMB lensing extraction.
We use the minimum variance estimator of Okamoto and Hu \cite{Oka03} to compute 
the lensing noise power spectrum for {\sc Planck} experimental configuration. \\
Taking into account the changes in the time-temperature relation of the primordial plasma and
the modification of the neutrino thermal potential, we  compute the projected gravitational 
potential power spectrum and its correlation with the temperature in the presence of DM sterile neutrino componet that together with the lensed CMB temperature and polarization anisotropy power spectra 
are used to constrain the cosmological parameters including sterile neutrino mass, from 
{\sc Planck}-like observations. \\
We show that the cosmological parameters are generally not biased when DM sterile neutrino is included. 
From this analysis we found a lower limit on DM sterile neutrino mass $m_{\nu_s}>2.08$ keV at 95\% CL, 
consistent with the lower mass limit obtained from the combined analysis of CMB, SDSS 3D power spectrum and SDSS Ly$\alpha$ forest power spectrum 
($m_{\nu_s}>1.7$ keV) \cite{AbaPrd}.\\
We conclude that, although the information that can be extracted is rather limited due to the high level of the lensing noise of  {\sc Planck} experiment, the weak lensing of CMB offers a valuable alternative to constrain the dark matter sterile neutrino mass.

\vspace{0.5cm}
{\bf Acknowledgements} \\

We thank to Oleg Ruchayskiy and  Alexey Boyarsky for the useful comments.\\
We acknowledge the use of the GRID computing system facility at
the Institute for Space Sciences Bucharest and the staff working there.

\end{document}